\documentclass[letter,longauth]{aa} 
\usepackage{graphicx}
\usepackage{txfonts}
\usepackage{natbib}
\bibpunct{(}{)}{;}{a}{}{,} 
\newcommand\micron{\mbox{$\mu$m}}%
\begin{document}
   \title{Determining dust temperatures and masses in the {\it Herschel}
    era: the importance of observations longward of 200 
    micron\thanks{{\it Herschel} is an ESA space observatory with science instruments provided
by European-led Principal Investigator consortia and with important participation from NASA.}}

   \titlerunning{{\it Herschel} dust temperatures and masses}

   \author{K. D. Gordon\inst{1}
    \and F. Galliano\inst{2}
    \and S. Hony\inst{2}
    \and J.-P. Bernard\inst{3}
    \and A. Bolatto\inst{4}
    \and C. Bot\inst{5}
    \and C. Engelbracht\inst{6}
    \and A. Hughes\inst{7}\fnmsep\inst{8}
    \and F. P. Israel\inst{9}
    \and F. Kemper\inst{10}
    \and S. Kim\inst{11}
    \and A. Li\inst{12}
    \and S. C. Madden\inst{2}
    \and M. Matsuura\inst{13}\fnmsep\inst{14}
    \and M. Meixner
      \inst{1}\fnmsep\inst{15}
    \and K. Misselt \inst{6}
    \and K. Okumura\inst{2}
    \and P. Panuzzo\inst{2}
    \and M. Rubio\inst{16}
    \and W. T. Reach\inst{17}\fnmsep\inst{18}
    \and J. Roman-Duval\inst{1}
    \and M. Sauvage\inst{2}
    \and R. Skibba\inst{6}
    \and A.G.G.M. Tielens\inst{9}}

\institute{Space Telescope Science Institute, 3700 San Martin Drive, Baltimore, MD 21218, USA
   \email{kgordon@stsci.edu}  
\and CEA, Laboratoire AIM, Irfu/SAp, Orme des Merisiers, F-91191 Gif-sur-Yvette, France
\and Centre d' \'{E}tude Spatiale des Rayonnements, CNRS, 9 av. du Colonel Roche, BP 4346, 31028 Toulouse, France
\and Department of Astronomy, University of Maryland. College Park, MD 20742, USA
\and Observatoire Astronomique de Strasbourg, 11, rue de l'universite, 67000 STRASBOURG, France
\and Steward Observatory, University of Arizona, 933 North Cherry Ave., Tucson, AZ 85721, USA
\and Centre for Supercomputing and Astrophysics, Swinburne University of Technology, Hawthorn VIC 3122, Australia
\and CSIRO Australia Telescope National Facility, PO Box 76, Epping NSW 1710, Australia
\and Sterrewacht Leiden, Leiden University, P.O. Box 9513, NL-2300 RA Leiden, The Netherlands
\and Jodrell Bank Centre for Astrophysics, Alan Turing Building, School of Physics \& Astronomy, University of Manchester, Oxford Road, Manchester M13 9PL, United Kingdom
\and Astronomy \& Space Science, Sejong University, 143-747, Seoul, South Korea
\and 314 Physics Building, Department of Physics and Astronomy, University of Missouri, Columbia, MO 65211, USA
\and Department of Physics and Astronomy, University College London, Gower Street, London WC1E 6BT, UK 
\and MSSL, University College London, Holmbury St. Mary, Dorking, Surrey RH5 6NT, U.K.
\and Visiting Scientist at Smithsonian Astrophysical Observatory, Harvard-CfA, 60 Garden St., Cambridge, MA, 02138, USA
\and Departamento de Astronomia, Universidad de Chile, Casilla 36-D, Santiago, CHILE
\and Spitzer Science Center, California Institute of Technology, MS 220-6, Pasadena, CA  91125, USA
\and Stratospheric Observatory for Infrared Astronomy, Universities Space Research Association, Mail Stop 211-3, Moffett Field, CA 94035 
}

  \abstract
{The properties of the dust grains (e.g., temperature and
mass) can be derived from fitting far-IR SEDs ($\geq$100~\micron).  Only
with SPIRE on {{\it Herschel}} has it been possible to get high spatial
resolution at 200 to 500~\micron\ that is beyond the peak
($\sim$160~\micron) of dust emission in most galaxies.}
{We investigate the differences in the fitted dust temperatures and
masses determined using only $<$200~\micron\ data and then also
including $>$200~\micron\ data (new SPIRE observations) to determine
how important having $>$200~\micron\ data is for deriving these dust
properties.}
{We fit the 100 to 350~\micron\ observations of the Large Magellanic
Cloud (LMC) point-by-point with a model that consists of a single
temperature and fixed emissivity law.  The data used are existing
observations at 100 and 160~\micron\ (from IRAS and {{\it Spitzer}}) and new
SPIRE observations of 1/4 of the LMC observed for the HERITAGE Key
Project as part of the {{\it Herschel}} Science Demonstration phase.}
{The dust temperatures and masses computed using only 100 and
160~\micron\ data can differ by up to 10\% and 36\%, respectively,
from those that also include the SPIRE 250 \& 350~\micron\ data.  We
find that an emissivity law proportional to $\lambda^{-1.5}$ minimizes
the 100--350~\micron\ fractional residuals.  We find that the emission
at 500~\micron\ is $\sim$10\% higher than expected from extrapolating
the fits made at shorter wavelengths.  We find the fractional
500~\micron\ excess is weakly anti-correlated with MIPS 24~\micron\
flux and the total gas surface density.  This argues against a flux
calibration error as the origin of the 500~\micron\ excess.  Our
results do not allow us to distinguish between a systematic variation
in the wavelength dependent emissivity law or a population of very
cold dust only detectable at $\lambda \geq 500~\micron$ for the origin
of the 500~\micron\ excess.}
   {}


   \maketitle
%

\section{Introduction}

Among nearby galaxies, the Large Magellanic Cloud (LMC) and Small
Magellanic Cloud (SMC) represent unique astrophysical laboratories for
interstellar medium (ISM) studies.  Both Clouds are relatively nearby,
the LMC at $\sim$50 kpc \citep{Schaefer08} and the SMC at $\sim$60 kpc
\citep{Hilditch05}, and provide ISM measurements that are relatively
unconfused by line-of-sight uncertainties when compared to the Milky
Way.  The two Clouds span an interesting metallicity range with the
LMC at $\sim$1/2 Z$_{\sun}$ \citep{Russell92} being above the
threshold of 1/3--1/4 Z$_{\sun}$ where the properties of the ISM
change as traced by the rapid reduction in the PAH dust mass fractions
and possible dust-to-gas ratios \citep{Draine07} and
the SMC at $\sim$1/5Z$_{\sun}$ \citep{Russell92} below this threshold.
Finally, the dust in the LMC and SMC shows strong variations in its
ultraviolet characteristics \citep{Gordon03}.

The HERschel Inventory of The Agents of Galaxy Evolution (HERITAGE) in
the Magellanic Clouds {\it Herschel} Key Project will map both Clouds using
the PACS/SPIRE Parallel observing mode providing observations at 100,
160, 250, 350, and 500~\micron\ \citep{Meixner10}.
The HERITAGE wavelength coverage (100--500~\micron) and spatial
resolution ($\sim$10~pc at 500~\micron) is well suited to measuring
the spatial variations of dust temperatures and masses.  The infrared
dust emission in most galaxies peaks between 100--200~\micron\
\citep{Dale05} and observations $>$200~\micron\ are important for
accurate dust temperature and masses \citep{Willmer09}.  Ground-based
submilimeter observations do provide the needed $>$200~\micron\
observations, but they have been seen to be in excess of that expected
from extrapolating fits to the $<$200~\micron\ data for sub-solar
metallicity galaxies \citep{Galliano05}.  This excess could be due to
very cold dust that only emits at submilimeter wavelengths or
variations in the wavelength dependent dust emissivity law
\citep{Reach95, Paradis09}.  As part of the Science Demonstration
Program (SDP), two HERITAGE AORs centered on the LMC were executed.
These observations are used in this paper to explore the impact SPIRE
observations have on the measurement of dust temperatures and masses
including the behavior of any submilimeter excess.


\section{Data}

\begin{figure*}
\resizebox{\hsize}{!}{\includegraphics{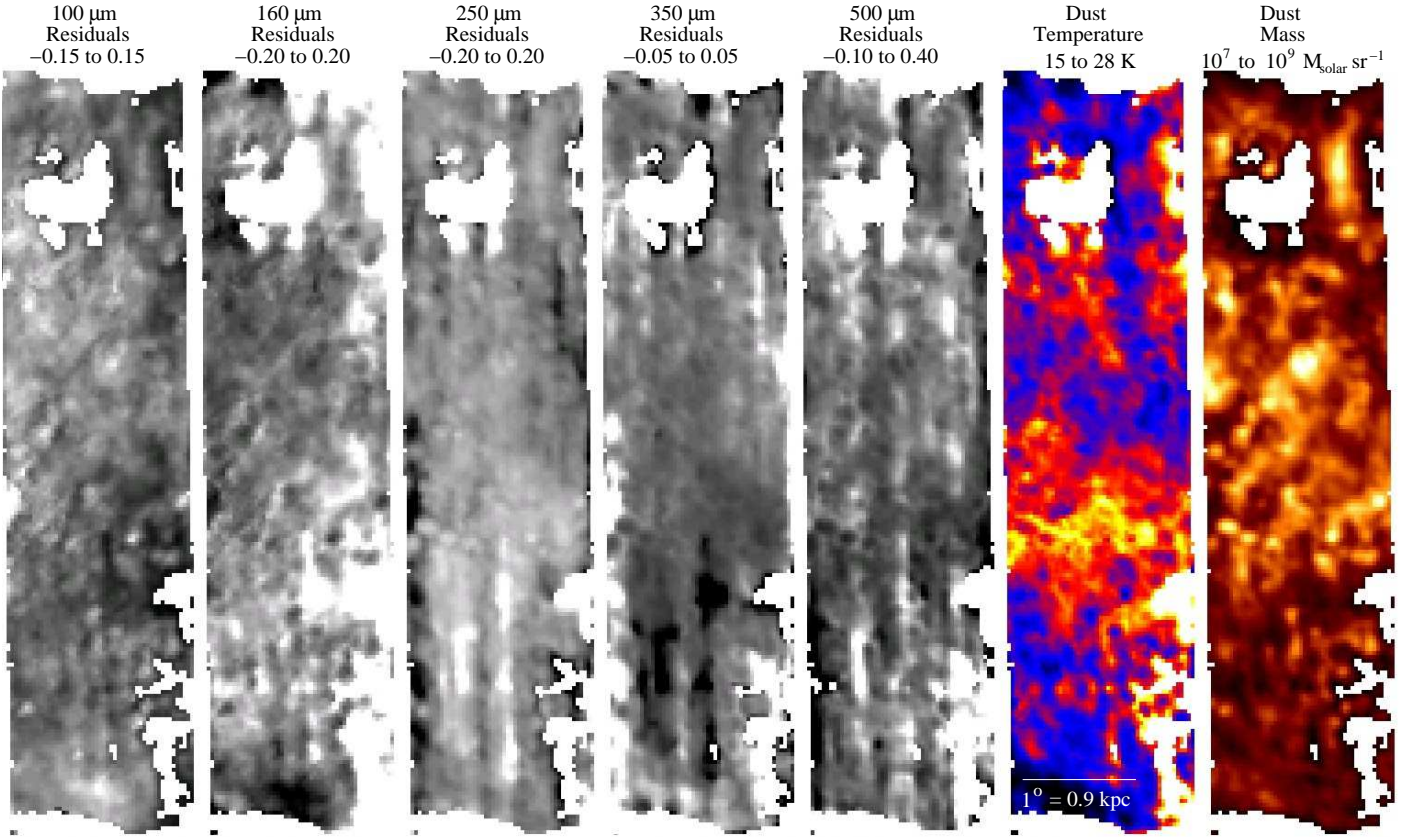}}
\caption{The best fit (assuming an emissivity law with $\beta = 1.5$)
dust temperature and mass images are shown along with the fractional
residual images at all 5 bands.  The resolution of all the images is
$4\farcm 3$.  Each of the fractional residual images shows the
difference in flux between the measured value and best fit model
divided by the best fit model.  They are separately linearly scaled
(black to white) to emphasize the structure and the scale ranges are
given at the top of each of the images.  The dust temperature image is
linearly scaled between 18 (blue) and 28~K (yellow).  The dust mass
image has a sqrt scaling between $10^7$ (black) and $10^9$ (yellow)
M$_\odot$/sr.  The vertical white and black streaks in the SPIRE
fractional residual images are caused by residual instrumental
signatures.}
\label{fig_resid_dust_temp_mass}
\end{figure*}

The observation and data reduction for the HERITAGE SDP data are given
in \citet{Meixner10}. For this paper, we use high
quality IRAS 100~\micron\ and MIPS 160~\micron\ observations instead
of the PACS observations which display large residual instrumental
signatures (expected to be eliminated with the full HERITAGE dataset).
We extracted the HERITAGE SDP region from the existing IRAS/IRIS
100~\micron\ \citep{Miville-Deschenes05} and MIPS 160~\micron\
\citep{Meixner06, Bernard08} mosaics.  We have used custom convolution
kernels created using the technique of
\citet{Gordon08} to convolve the images to a common resolution of
$4\farcm 3$ of the IRAS 100~\micron\ data.  We also created a 2nd set of
images (excluding the IRAS 100~\micron\ data) at the common 
resolution of the SPIRE 500~\micron\ and MIPS 160~\micron\ data of
$\sim$38$\arcsec$.

Emission from Milky Way (MW) foreground cirrus clouds contributes to
the far-IR emission seen in the LMC.  We use the HI column density map
created by integrating the MW velocities in the full HI cube
\citep{Staveley-Smith03} to correct all the images for the MW infrared cirrus
emission.  The HI column densities were transformed to IR surface
brightnesses using the model of the MW emission used by
\citet{Bernard08}.  Finally, any residual emission was removed
by fitting a gradient across the SDP region using the regions in the
strip beyond the IR edge of the LMC.


\section{Results}

For each point in the image, we determined the dust temperature by
fitting the observed far-IR SED to a modified black body of the form
\begin{equation}
F_{\nu} \propto \lambda^{-\beta} B_{\nu}(T_{dust})
\end{equation}
The dust mass is computed from the measured 160~\micron\ flux
($F_{160}$), at each point, using
\begin{equation}
M_{dust} = \frac{4}{3} \frac{a \rho
d^2}{Q_{em}(160)}\frac{F_{160}}{B_{\nu}(T_{dust})}
\end{equation}
where $a = 0.1~\micron$ is the grain radius, the grains are assumed to
be spherical silicate grains with a density $\rho = 3$~g~cm$^{-3}$, d
= 50~kpc is the LMC distance, and $Q_{em}(160) = 5.5\times 10^{-4}$
\citep{Laor93}.  This method is fairly standard and while other more
sophisticated fitting methods exist \citep{Draine07, Galliano08b,
Paradis09LMC}, this simple model allows us to probe the effects of
adding $>$200~\micron\ data to the fits with fewest assumptions.  We
restrict our fits to using only data $\geq$$100~\micron$ as
observations at shorter wavelengths likely include non-equilibrium
dust grain emission (transient heating).  The data points used in the
fits were weighted by the uncertainties (i.e., $1/\sigma^2$).  The
main uncertainties on the measurements are the calibration and
background noise uncertainties and we sum them in quadrature.  The
calibration errors are assumed to be approximately the same at all
bands at around 15\% \citep{Stansberry07MIPS160, Swinyard10}.  Data
within 1$\sigma$ of the background are not used in the fits. 

\subsection{Dust temperatures and masses}

The best fit dust temperature and mass values
were determined by for fits using only the pre-{\it Herschel} data
(IRAS 100~\micron\ and MIPS 160~\micron) and fits including
the {\it Herschel} SPIRE data (IRAS 100~\micron, MIPS 160~\micron, and SPIRE
250/350~\micron).  Given the inclusion of the IRAS 100~\micron, we
used the $4\farcm 3$ resolution images.  The SPIRE 500~\micron\ data
are not included in these fits as it is usually systematically high
(see Sec.~\ref{sec_500excess} and \citet{Meixner10}) and including it in the fits only
causes the residuals at the other wavelengths to increase without
significantly improving the fit.  The value of $\beta$ used in the
fits was set to 1, 1.5, or 2 as this range encompasses realistic dust
grains (amorphous to crystalline grains) and is what has been used in
the past \citep{Dunne00}.  The dust temperature and
mass maps and fractional residual images for the $\beta = 1.5$ case
are shown in Fig.~\ref{fig_resid_dust_temp_mass}.

The differences between the best fit dust temperatures and masses (with and
without the SPIRE data) depends on the value of $\beta$ used.  For
$\beta = 2$, the with SPIRE to without SPIRE temperature ratio is
$0.97 \pm 0.06$ and mass ratio is $1.19 \pm 0.31$.  For $\beta = 1.5$,
the with/without SPIRE temperature ratio is $1.02 \pm 0.07$ and mass
ratio is $0.96 \pm 0.25$.  For $\beta = 1$, the with/without
temperature ratio is $1.08 \pm 0.08$ and mass ratio is $0.77 \pm
0.20$.  Thus, the inclusion of $>200~\micron$ data in the fits can
change the derived dust temperature by up to 8\% and mass by up to
23\% depending on the assumed value of $\beta$.

\begin{figure}
\resizebox{\hsize}{!}{\includegraphics{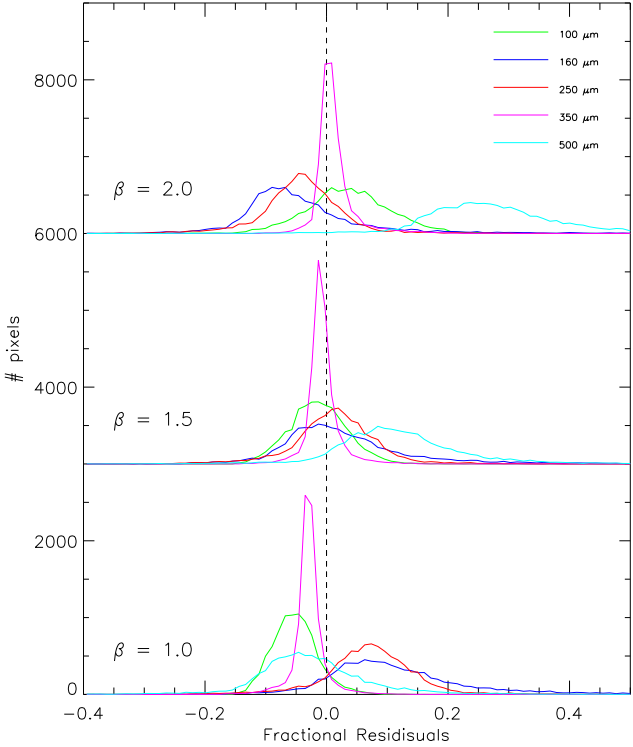}}
\caption{The histograms of the fractional residuals at different
wavelengths are shown for $\beta = $1.0, 1.5, and 2.0.  The fractional
residual is the difference in flux at each wavelength between the
measured and best fit model divided by the best fit model.  The $\beta
= 1.5$ and 2.0 histograms are offset by 3000 and 6000 pixels,
respectively.  The dashed vertical line indicates zero fractional
residual.  The strongly peaked 350~\micron\ histogram is simply a
result of the relative weighting of different wavelengths in the fit.
For example, using equal weights produces more equally peaked
histograms between the different wavelengths.}
\label{fig_allbeta}
\end{figure}

Prior to the {\it Herschel} observations, it was not possible to constrain
the best value of $\beta$ given that there were only two infrared maps
of the LMC with $\lambda \geq 100~\micron$.  With the {\it Herschel}
observations, the behavior of the residuals as a function of $\beta$
can be used to determine the optimal $\beta$ value.  Histograms of the
fractional residuals at different $\beta$ values are shown in
Fig.~\ref{fig_allbeta}.  A value of $\beta = 1.5$ clearly minimizes
the fractional residuals at all wavelengths with most of the pixels
having residuals of less 10\% at all wavelengths except 500~\micron.
This result implies that either the characteristics of the dust grains
are intermediate between the two extremes or that a more complex dust
emission model including a distribution of dust temperatures and grain
sizes is needed \citep{Draine07, Paradis09LMC}.  Assuming a $\beta =
2.0$ for the pre-{\it Herschel} fits (a common assumption) and using the
best fit $\beta = 1.5$ for the fits including the SPIRE data, we find
the with/without temperature ratio is $1.12 \pm 0.07$ and mass ratio
is $0.64 \pm 0.16$.  This decrease in dust masses reduces the
magnitude of the ``FIR excess'' found by \citet{Bernard08} for the
LMC.  \citet{Roman-Duval10} explore this issue in
detail for two specific LMC molecular clouds.

 
\subsection{500 \micron\ excess}
\label{sec_500excess}

In the previous section, we have not included the 500~\micron\
observations in the analysis as it was seen not to improve the quality
of the fits and previous studies \citep{Galliano05, Galametz09} have
observed submm fluxes in excess of that expected from fits to the
far-IR fluxes.  At $4\farcm 3$ resolution, the average fractional
500~\micron\ fit residual is 0.25, 0.10, and -0.05 for $\beta$ values
of 2, 1.5, and 1 (Fig.~\ref{fig_allbeta}).  As a $\beta = 1.5$ is
strongly favored as it minimizes the residuals at all other
wavelengths, we find a 500~\micron\ excess of approximately 10\%.  We
find the same level of 500~\micron\ excess for fits done at both the
$4\farcm 3$ and $38\arcsec$ resolutions.

\begin{figure*}
\resizebox{\hsize}{!}{\includegraphics{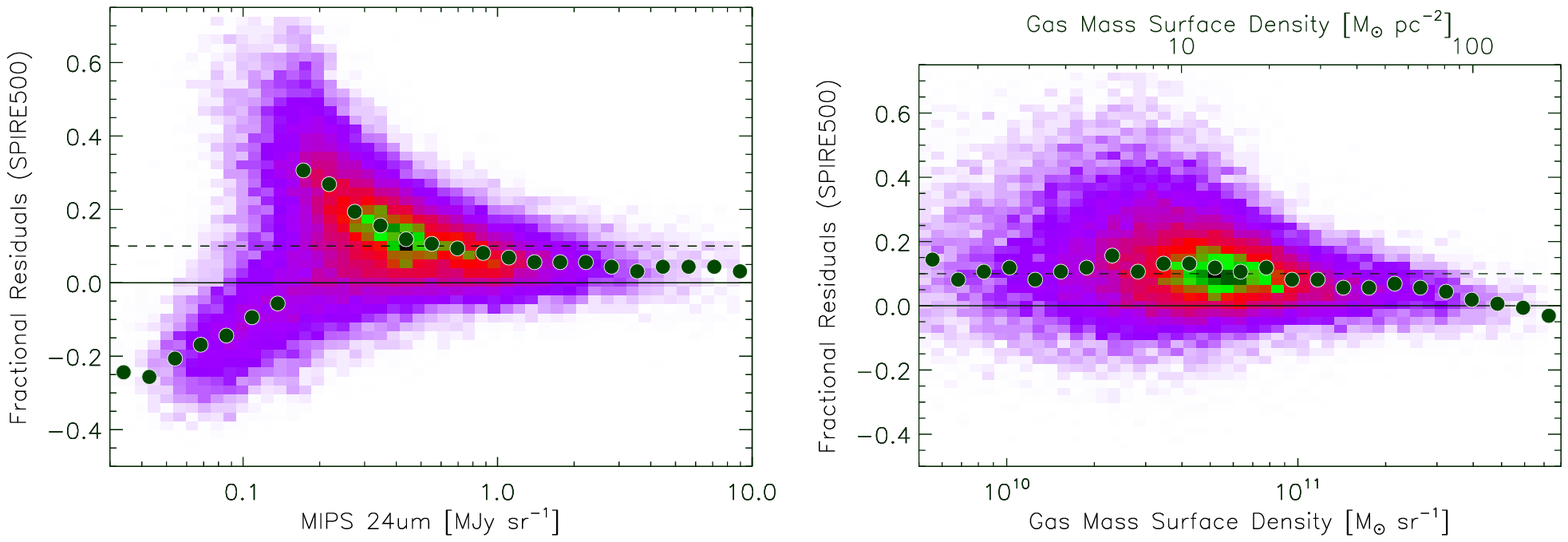}}
\caption{The 500~\micron\ excess is plotted versus MIPS 24~\micron\
flux and total gas mass surface density using the $38\arcsec$
resolution images.  The total gas mass is derived from HI and CO
observations \citep{Kim03, Fukui08} assuming $X_{CO} = 7\times
10^{20}$ cm$^{-2}$ (K km s$^{-1}$)$^{-1}$ \citep{Fukui08} and an
overall 36\% increase in the gas mass to account for the associated He
gas.  The density of points is coded from low to high as
purple-red-green-black.  The solid and dashed black lines give the
zero and 10\% excess values, respectively.  The solid points give the
values of the mode in equally spaced logarithmic bins.  On average,
the residuals are markedly negative at low MIPS 24~\micron\ fluxes
indicating that below these fluxes, the residuals are affected by
background subtraction or fitting errors.}
\label{fig_excess_vs_24_temp}
\end{figure*}

There are four possible origins of the 500~\micron\ excess: 1)
systematics due to our assumptions on our fitting, 2) a flux
calibration error, 3) variations in the wavelength dependent
emissivity law
\citep{Reach95, Agladze96}, and 4) very cold dust that mostly emits at $\geq
500~\micron$ \citep{Finkbeiner99, Galliano05}.  Whatever the
the physical process responsible for the 500~\micron\ excess, the 
HERITAGE SDP SPIRE data of the
LMC allow us to probe the origin of the 500~\micron\ excess at high
spatial resolution in an external galaxy for the first time.  We
tested the systematics of our fitting algorithm and searched for
correlations of the 500~\micron\ excess with different tracers of the
ISM conditions (dust temperature, dust mass, HI mass, and MIPS
24~\micron\ flux) in an attempt to determine the origin of the
500~\micron\ excess.  The two strongest correlations are seen for MIPS
24~\micron\ flux (probing the ISM conditions for small grains) and the
total gas mass (probing the ISM density) are are shown in
Fig.~\ref{fig_excess_vs_24_temp}.

To test 1), we performed Monte Carlo simulations where the
observations were simulated both with and without an excess at 500~\micron\
and with different $\beta$ values.  These simulations were fit with
varying $\beta$ laws and realistic uncertainties.  A 500~\micron\
excess was found in the simulations if it was part of the
simulation or if the fitting $\beta$ was smaller than the simulation
$\beta$.  Given that we empirically determine $\beta$ from the $<
500~\micron$ data, our conclusion is that the excess we find is not a
result of our fitting method.

For 2), it is possible that there is a systematic 500~\micron\ flux
calibration error on the order of 10\%.  The official maximal possible
flux calibration error for SPIRE is 15\% at any wavelength
\citep{Griffin10}.  Given that we are including the SPIRE 250 and
350~\micron\ measurements in our fitting, the 500~\micron\ flux
calibration error would have to be relative to the other two SPIRE
bands and so is likely smaller than 15\%.  In addition, we would
expect to see no correlation between the excess and ISM condition
tracers, yet we see weak correlations
(Fig.~\ref{fig_excess_vs_24_temp}).

For 3), a wavelength dependent increase in the dust emissivity law at
500~\micron\ on the order of 10\% is possible \citep{Paradis09}.  This
variation may be attributed to the dust grains
amorphous/crystalline nature, size distribution, temperature, and
material \citep{Henning95, Meny07}.  For example, if the 500~\micron\
excess is due to small grains having a different $\beta$ than large
grains, we would expect the excess to be correlated with the MIPS
24~\micron\ emission
\citep{Reach95}.  Yet the excess is weakly anti-correlated with MIPS
24~\micron\ flux (Fig.~\ref{fig_excess_vs_24_temp}).

For 4), very cold dust that only emits at $\geq$500~\micron\ is
physically possible.  The very coldest dust would necessarily be the
dust that is best shielded and, thus, we would then expect the excess
to be strongest in the highest density regions and regions with the
lowest radiation fields.  Figure~\ref{fig_excess_vs_24_temp} gives a
conflicting answer as we see the largest excesses in the faintest
24~\micron\ regions (as expected) and least dense regions (not as
expected).


\section{Conclusions}

We investigate the importance of $>$200~\micron\ data in determining
dust temperatures and masses using new {\it Herschel} SPIRE observations of
the LMC (taken for the HERITAGE Key Project as part of the
{\it Herschel} Science Demonstration phase) combined with existing IRAS
100~\micron\ and {\it Spitzer} MIPS 160~\micron\ images.  We fit the
observations with a model consisting of dust emitting as a single
temperature blackbody modified with an emissivity law proportional to
$\lambda^{-\beta}$.  For fixed values of $\beta$, fits using only the
100--160~\micron\ data give dust temperatures and masses that are on
average up to 8\% and 23\% different from fits using the same $\beta$
and the 100--350~\micron\ data.  The new SPIRE observations allowed us
to determine that $\beta = 1.5$ minimizes the residuals from 100 to
350~\micron.  Using a $\beta = 2.0$ for the 100--160~\micron\ and a
$\beta = 1.5$ for the 100--350~\micron\ fits results in an increase of
10\% for the dust temperature and a decrease in the dust mass by 36\%.

On average, there is a fractional excess at 500~\micron\ of
$\sim$10\%.  The origin of the fractional excess is unlikely to be due
to our fitting algorithm or a flux calibration error, but it could be
due to either very cold dust that emits only $\geq$500~\micron\ or a
variation in the wavelength dependent change in the dust emissivity.
Planned HERITAGE observations of the LMC and SMC will allow for a more
detailed investigation of including $> 200~\micron$ data (mainly the
500~\micron\ excess) due to better quality PACS and SPIRE images
(optimized observations and cross-scans).

\begin{acknowledgements}
We acknowledge financial support from the NASA {\it Herschel} Science
Center, JPL contracts \# 1381522 \& 1381650.  M.R.\ is supported by
FONDECYT No1080335 and FONDAP No15010003. We thank the contributions
and support from the European Space Agency (ESA), the PACS and SPIRE
teams, the {\it Herschel} Science Center and the NASA {\it Herschel} Science Center (esp.\
A.\ Barbar and K.\ Xu) and the PACS and SPIRE data center at
CEA-Saclay, without which none of this work would be possible.  We
thank the referee, Thomas Henning, for suggestions that improved the
clarity of this paper.
\end{acknowledgements}

\end{document}